\documentstyle[preprint,revtex,eqsecnum]{aps}
\begin{document}
\preprint{IFT-P.055/93}
\preprint{September 1993}
\preprint{hep-ph/9309210}
\begin{title}
{\Large \center  Constraints on spin-$\frac{3}{2}$ and excited
spin-$\frac{1}{2}$ fermions  \\
coming from the leptonic $Z^0$-partial width \\}
\end{title}
\author{ J.C. Montero and V. Pleitez}
\begin{instit}
Instituto de F\'\i sica Te\'orica\\
Universidade Estadual Paulista--UNESP, Rua Pamplona, 145\\
01405-900--S\~ao Paulo, SP, Brasil.
\end{instit}
\begin{abstract}
We consider effective interactions among excited
spin--$\frac{1}{2}$ and spin--$\frac{3}{2}$  leptons with the usual
ones. Assuming that these new
leptons are lighter
than the $Z^0$ we will study the constraints on their masses
and compositeness scale coming from the leptonic $Z^0$ partial width.
\end{abstract}
\pacs{PACS numbers: 14.60.Jj, 14.80.-j }
\newpage
One way to study possible new physics beyond the standard model,
which is based on $SU(3)_c\otimes SU(2)_L\otimes U(1)_Y$ gauge
symmetry~\cite{mp}, is to assume new
particles and interactions and compare the results of this
new physics with the experimental data, say, with
$a_\mu=\frac{1}{2}(g_\mu-2)$ measurements or in cross sections and
angular distributions in the forthcoming accelerators. This new
particles and interactions would be discover not very far from the
Fermi scale~\cite{wb,fs}.

If the known fermions are composite, they will have
additional contact interactions. A consequence of composite models
is the existence of excited fermions and higher spins particles, for
instance excited fermions with spin--$\frac{3}{2}$. Usually in this sort of
picture it is introduced a compositeness scale $\Lambda_{comp}$
related to the extension of the composite particle. This scale also
define two different regimes for the interactions in the following
sense: for scales below $\Lambda_{comp}$ we get the regime of bound
states and the interactions appear as effective interactions whereas
for scales above $\Lambda_{comp}$ the composite particles appear as a
free system of constituents with an underlying, up to now unknown,
dynamics.

{}From the theoretical point of view there is no universal agreement on
which scale the lack of elementarity becomes operative. Possibilities
run from $\Lambda_{comp}\sim 1\,\mbox{TeV}$ to
$\Lambda_{comp}\approx M_{Planck}\sim
10^{19}\,\mbox{GeV}$. To decide among these options require more
experimental data on the predictions of the standard model, in such a
way that we could identify the contributions of new physics, if any.

Usually new leptons are assumed to have a large mass, say
150--200 GeV. Next it is possible to look for their effects
in colliders as $e^-p$~\cite{simoes}.
In general, it is consider that the masses of the excited fermions are of
the same order of magnitude of the compositeness scale. However, in
principle, some excitations could have masses even comparable to those
of the vector bosons.

Here, we will assume that in fact,
excited or higher spin leptons may have masses below $M_Z$ in such
a way that they contribute to the $Z^0$ width, being produced
accompanying a usual lepton. Hence, we will consider  the decay
$Z^0\to \Psi l$, with $\Psi$
denoting a new lepton and $l$ a usual one ($l=e,\mu,\tau$).

The possibility of the existence of a composite structure for, up to
now, elementary particle has been confronted with the $a_\mu$
measurements~\cite{drell,r1}. Also tests of composite models coming
from $Z^0$ decays modes have already
been considered~\cite{r2}. For example, decay rates and branching
ratios for production cross sections in $p\bar p$, $pp$ and $e^+e^-$
and angular distributions have been studied~\cite{ee,kuhn}. It is important to
identify new observables which could be sensitive to the excited
and/or higher spin fermions. Since there
exist at present very precise measurements of the total and partial
$Z^0$ decay width, it is also possible, to use this $Z$-pole
observable to constrain new physics. For instance, this has been done
in the simple extension of the standard model which include beside
the usual sequential leptons, a singlet right-handed neutral
field~\cite{tau3}.

Here we will use this method to constrain the effective
interactions involving excited spin--$\frac{1}{2}$ and
spin--$\frac{3}{2}$ fields. These
interactions are not $SU(2)$ gauge invariant as is usually to be
consider nowadays~\cite{gavela} since we do not want to exclude models
where the $Z^0$ and $W^\pm$ are composite too. In this case, the
standard model $SU(2)_L\otimes U(1)_Y$ gauge theory is replaced by an
effective global theory in which the $W$ and $Z^0$, being not
elementary,  may not couple as in the standard model.

We will get bound on $\Lambda_{comp}$ and the masses of the new
particles by assuming that the experimental data of the $Z^0$ width
is given by the standard model predictions so that the new physics has
room only in the experimental uncertainties.

There are many possible effective interactions connecting the
usual and excited leptons. Here we will consider only three of them
just to illustrate how the experimental data of the
$Z^0 \to l \bar l$ partial width constrains the scale of compositeness and the
new lepton masses. In the following, for economy, we will use
$\Lambda$ and $m^*$ to denote the compositeness scale and the new
lepton masses respectively for both the excited spin-$\frac{1}{2}$
fermion ($\psi^*$)and the spin-$\frac{3}{2}$ one ($\chi$).
The mass of the usual leptons ($l$)
will be denoted by $m_l$.

\noindent {\bf Exited Spin-$\frac{1}{2}$ Leptons}. We will assume
that the excited leptons with spin--$\frac{1}{2}$,
$\Psi=\psi^*$, couples to the usual leptons through the
effective interaction
\begin{equation}
{\cal L}_{\psi^*}=\frac{1}{2\Lambda}\bar\psi^*\sigma_{\mu\nu}
(c+d\gamma_5)l Z^{\mu\nu}+H.c.
\label{int1}
\end{equation}
By $CP$ invariance $c$ and $d$ have to be real and if their origin is
some mixing in the leptonic sector these constants must be
normalized to unity, $c^2+d^2=1$. Throughout this work we will assume
this is the case but we must remember that if the origin of $c$ and
$d$ is a mixing in the gauge boson sector these coefficients do not
need to be normalized to unity as occurs in the neutral currents in
the standard electroweak model. We will also use, the usual
normalization $g^2/4\pi=1$ with $g^2$ a dimensionless constant.

\noindent {\bf Spin-$\frac{3}{2}$ Leptons}. Here we will consider two
effective interactions of a spin--$\frac{3}{2}$ particle, $\Psi=\chi$,
with the usual leptons,
\begin{equation}
{\cal L}_{\chi}=g\bar\chi^\mu(c+d\gamma_5)l Z_\mu+H.c.,
\label{int2}
\end{equation}
with $g$ a dimensionless constant. As in this case the
strength of the interaction does not depend  on the scale $\Lambda$,
we will not consider the $g^2/4\pi=1$ normalization.

We will also consider the interaction
\begin{equation}
{\cal L}'_{\chi}=\frac{1}{\Lambda}\bar\chi^\mu\gamma^\nu
(c+d\gamma_5)l Z_{\mu\nu}+H.c.
\label{int3}
\end{equation}

\noindent{\bf $Z^0$ Partial Width}. As we said before, it is possible
that the excited leptons have a mass smaller than the
compositeness scale. If they are sufficiently light they may appear
as a decay product in $Z^0$ decay~\cite{kuhn}. In this case the
$Z^0$-partial width is
given by
\begin{equation}
\Gamma(Z\rightarrow \Psi l)=\frac{1}{2\pi}\frac{1}{(2J+1)M^2_Z}
\left[\frac{(M^2_Z-m^2_*+m^2_l)^2}{4M^2_Z}-m^2_l\right]^{\frac{1}{2}}
F_\Psi(M^2_Z,m^2_l,m^2_*),
\label{z0}
\end{equation}
where $\Psi$ denotes any of the new particles
 we are considering here. $J$ is the spin of the decaying
particle i.e., $J=1$ in this case.

For the case of excited spin--$\frac{1}{2}$ leptons
we have, for the interaction given by Eq.~(\ref{int1})
\begin{eqnarray}
F_{\psi^*}&=&\frac{(c^2+d^2)}{\Lambda^{2}}\left[4\left(M^2_Z
-(m^2_*-m^2_l)^2\right)
-2M^2_Z(M^2_Z-m^2_*-m^2_l)\right]\nonumber \\ \mbox{} & &
 +\frac{12}{\Lambda^{2}}(c^2-d^2)M^2_Zm^*m_l.
\label{1}
\end{eqnarray}
For the spin--$\frac{3}{2}$ we have,
\begin{eqnarray}
F_{\chi}&=&\frac{8}{3}g^2\left(2+\frac{(M^2_Z+m^2_*-m^2_l)^2}{4m^2_*M^2_Z}
\right)[\frac{c^2+d^2}{4}(M^2_Z-m^2_*-m^2_l)
\nonumber \\ \mbox{} & &
\left.-m^*m_l(c^2-d^2)]\right.,
\label{2}
\end{eqnarray}
assuming the interaction given by Eq.~(\ref{int2}) and
\begin{eqnarray}
F'_{\chi}&=&\frac{2}{3}\frac{c^2+d^2}{\Lambda^2 m^3_*}
[3m^*m^4_lM^2_Z-m^*m^6_l-5m^3_*m^4_l\nonumber \\ \mbox{}
& & +13m^5_*m^2_l + 8m^3_*m^2_lM^2_Z-3m^*m^2_lM^2_Z\nonumber \\
\mbox{} & &-7m^7_*+9m^5_*M^2_Z-3m^3_*M^4_Z+m^*M^6_Z]
\nonumber \\ \mbox{} & &-24\frac{c^2-d^2}{\Lambda^2}m^*m_lM^2_Z.
\label{3}
\end{eqnarray}
from the interaction of Eq.~(\ref{int3}).

Note that in Eqs.(\ref{1})--(\ref{3}) is not possible to distinguish
between pure left- or pure right-handed interactions.
As usual to avoid a conflict with the
$g-2$ measurements it
is demanded that only either left-handed or right-handed
leptons couple to the excited ones. Hence we will use in the following
$c=d=\frac{1}{\sqrt{2}}$.

We will consider the two parameters $\Lambda$ and
$m^*$ as independent. Next, we will consider the constraints
on these parameters coming from the leptonic $Z^0$-partial width
for each effective interaction Eq.(\ref{int1})--Eq.(\ref{int3}).

The effect of excited leptons or
quarks in the $Z^0$ width has already
been studied~\cite{kuhn}, but only indirectly since the experimental
parameter was the cross sections of processes as proton-(anti)proton
and $e^+e^-$ collisions. Recently, measurements of the $Z^0$ partial
widths have been made very accurately~\cite{pdg}.

Here we will consider in particular the leptonic $Z^0$ partial
width as the measured parameter sensible, in principle, to the
effects of compositeness. In particular we will use the ratio
\begin{equation}
R_l=\frac{\Gamma_l}{\Gamma}\times 100 \%,
\label{ratio}
\end{equation}
where $\Gamma_l$ is the $Z^0\to l\bar l$ width
and $\Gamma$ is the total width.

We leave unrelated the values of $\Lambda$  and
$m^*$,
but kinematically $m^*\leq M_Z-m_l$. Experimental data of the $Z^0$
leptonic partial width will give the allowed region for both
parameters. Here we have used $R_e=3.345\pm0.025\,\%$ and
$R_\tau=3.320\pm 0.04\,\%$~\cite{pdg}.

In Fig.~1 we show the allowed and forbidden regions for $\Lambda$
and $m^*$ entering in Eq.~(\ref{int1}). In Figs.~2 and 3 we have done
the same for the parameters for the interactions given in
Eqs.~(\ref{int2}) and (\ref{int3}). There is a high sensitivity to
changes in the parameters ($g,m^*$) and ($\Lambda,m^*$) for the
interactions given by Eq.~(\ref{int2}) and Eq.~(\ref{int3}) as can be
seen in Fig.~2 and 3 respectively. On de other hand the
interaction (1) is not too sensitive to variations on the
($\Lambda,m^*$) parameters (see Fig.~1).

Notice that with the interaction (1) (Fig.1) for a $\Lambda$ of
about $5-9\,TeV$ the new particles can have arbitrary small masses.
Notwithstanding with interaction (3) (Fig.3) for the same $\Lambda$
values only particles with masses somewhat greater than
$\sim 35\,GeV$ are allowed. With the interaction (2) for arbitrarily
small values of the coupling constant $g$ it is also possible for
the new particles be arbitrarily light (Fig.~2).

We would like to stress that it is not straightforward to compare the
constraints we have obtained in this work with others existing in
the literature. In the later ones usually it is assumed at beginning
that $\Lambda = m^*$ \cite{r2}. The case $\Lambda >>m^*$ is usually
considered for the flavor changing transitions like
$\mu \to e\gamma$ \cite{r2}.

\acknowledgements

We would like to thank the
Con\-se\-lho Na\-cio\-nal de De\-sen\-vol\-vi\-men\-to Cien\-t\'\i
\-fi\-co e Tec\-no\-l\'o\-gi\-co (CNPq) for full (JCM) and partial
(VP) financial support.

\begin{center}
FIGURE CAPTION
\end{center}

\noindent
Fig.~1. Exited spin--$\frac{1}{2}$ fermion  coupled as
in Eq.~(\ref{int1}). The allowed and forbidden regions in the
($\Lambda,m^*$) plane delimited by the curves considering $m_l=m_e$
(continuous) and $m_l=m_\tau$ (dashed) for $1\sigma$ and $2\sigma$
experimental standard deviations on $R_l$ coming from Eq.~(\ref{1}) with
$c=d=\frac{1}{\sqrt2}$.\\

\noindent
Fig.~2. Spin--$\frac{3}{2}$ fermion coupled as in
Eq.~(\ref{int2}). The allowed and forbidden regions in the
($g,m^*$) plane delimited by the curves considering
$m_l=m_e$ (continuous) and $m_l=m_\tau$ (dashed) for $1\sigma$ and
$2\sigma$ experimental standard deviations on $R_l$ coming from
Eq.~(\ref{2}) with $c=d=\frac{1}{\sqrt2}$.\\

\noindent
Fig.~3. Spin--$\frac{3}{2}$ fermion coupled as in
Eq.~(\ref{int3}).  The allowed and forbidden regions in the
($\Lambda,m^*$) plane delimited by the curves considering
$m_l=m_e$ (continuous) and $m_l=m_\tau$ (dashed) for $1\sigma$ and
$2\sigma$ experimental standard deviations on $R_l$ coming from
Eq.~(\ref{3}) with $c=d=\frac{1}{\sqrt2}$.\\

\end{document}